\documentclass{aa}  

\usepackage{graphicx}
\usepackage{comment}
\usepackage{txfonts}
\usepackage[hidelinks]{hyperref}
\hypersetup{
colorlinks=true,
linkcolor=blue,
filecolor=blue,
citecolor = blue,
urlcolor=cyan,
}
\newcommand{\src}{3C~58{ }}
\begin{document}

   \title{A polarized view of the young Pulsar Wind Nebula \src { }with IXPE}
\titlerunning{IXPE view of 3C 58}    
   %\subtitle{bla bla bla...}

   \author{
    N.~Bucciantini\inst{\ref{inst1},\ref{inst2},\ref{inst3}}
    \and
    J.~Wong\inst{\ref{inst4}}  % move her here??; she was in 5th
    \and
    R.W.~Romani\inst{\ref{inst4}}
    \and
    F.~Xie\inst{\ref{inst5}}
    \and
    C.-Y.~Ng\inst{\ref{inst6}}
    \and
    S.~Silvestri\inst{\ref{inst7}}
    \and
    N.~Di~Lalla\inst{\ref{inst4}}
    \and
    %Y.-J.~Yang\inst{\ref{inst6},\ref{inst8},\ref{inst9},\ref{inst10}}
    Y.-J.~Yang\inst{\ref{inst10},\ref{inst8}}
    \and
    S.~Zhang\inst{\ref{inst6}}
    \and
    P.~Slane\inst{\ref{inst11}}
    \and
    W.-T.~Ye\inst{\ref{inst12},\ref{inst13}}
    \and
    M.~Pilia\inst{\ref{inst14}}
    \and
    N.~Omodei\inst{\ref{inst4}}
    \and
    M.~Negro\inst{\ref{inst15}}
          }

   \institute{INAF - Osservatorio Astrofisico di Arcetri, Largo Enrico Fermi 5, 50125, Firenze, Italy. \email{niccolo.bucciantini@inaf.it}\label{inst1}
   \and
   Dipartimento di Fisica e Astronomia, Universit\`{a} degli Studi di Firenze, Via Sansone 1, 50019, Sesto Fiorentino, Italy\label{inst2}
   \and
   Istituto Nazionale di Fisica Nucleare, Sezione di Firenze, Via Sansone 1, 50019, Sesto Fiorentino, Italy\label{inst3}
   \and
   Department of Physics and Kavli Institute for Particle Astrophysics and Cosmology, Stanford University, Stanford, California 94305, USA\label{inst4}
   \and
   Guangxi Key Laboratory for Relativistic Astrophysics, School of Physical Science and Technology, Guangxi University, Nanning 530004, China\label{inst5}
   \and
   Department of Physics, The University of Hong Kong, Pokfulam, Hong Kong \label{inst6}
   \and
   Istituto Nazionale di Fisica Nucleare, Sezione di Pisa, Largo B. Pontecorvo 3, 56127 Pisa, Italy\label{inst7}
   \and
   Graduate Institute of Astronomy, National Central University, 300 Zhongda Road, Zhongli, Taoyuan 32001, Taiwan\label{inst10}
   \and
   Laboratory for Space Research, The University of Hong Kong, Cyberport 4, Hong Kong \label{inst8}
%   \and
%   Department of Physics, National Cheng Kung University, University Road, Tainan, Taiwan\label{inst9}
   \and
   Center for Astrophysics | Harvard \& Smithsonian, 60 Garden Street, Cambridge, MA 02138, USA\label{inst11}
   \and
   Key Laboratory of Particle Astrophysics, Institute of High Energy Physics, Chinese Academy of Sciences, Beijing 100049, China\label{inst12}
   \and
   University of Chinese Academy of Sciences, Chinese Academy of Sciences, Beijing 100049, China\label{inst13}
   \and 
   INAF Osservatorio Astronomico di Cagliari, Via della Scienza 5, 09047 Selargius (CA), Italy\label{inst14}
   \and
   Department of Physics and Astronomy, Louisiana State University, Baton Rouge, LA 70803, USA\label{inst15}}
   \date{Received September X, XXXX; accepted March X, XXXX}

% \abstract{}{}{}{}{} 
% 5 {} token are mandatory
 
\abstract{Pulsar Wind nebulae (PWNe), are among the most efficient particle accelerators in the Universe, however understanding the physical conditions and the magnetic geometry in their inner region has always proved elusive. X-ray polarization provides now a unique opportunity to investigate the magnetic field structure and turbulence properties close to where high energy particles are accelerated. Here we report on the recent X-ray polarization measurement of the PWN \src by the International X-ray Polarimeter Explorer (IXPE). \src is a young system displaying a characteristic jet-torus structure which, unlike other PWNe, is seen almost edge on. This nebula shows a high level of integrated polarization $\sim 22\%$ at an angle $\sim 97^\circ$, with an implied magnetic field oriented parallel to the major axis of the inner torus, suggesting a toroidal magnetic geometry with little turbulence in the interior, and an intrinsic level of polarization possibly approaching the theoretical limit for synchrotron emission. No significant detection of a polarized signal from the associated pulsar was found. These results confirm that the internal structure of young PWNe is far less turbulent than previously predicted, and at odds with multidimensional numerical simulations.}


   \maketitle
%
%-------------------------------------------------------------------

\section{Introduction}
3C 58 (G130.7+3.1) is a young PWN, which shares many similarities with other systems like the Crab nebula, Vela and MSH 15-5\textit{2}. It has been detected throughout the electromagnetic spectrum from radio to X-rays and TeV $\gamma$-rays \citep{Green_Baker+75a,Becker_Helfand+82a,Davelaar_Smith+86a,Torii_Slane+00a,Bocchino_Warwick+01a,Slane_Helfand+04a,Slane_Helfand+08a,Fesen_Rudie+08a,Shibanov_Lundqvist+08a,Aleksic_Ansoldi+14a,Li+2018A,An19a}. Its apparent size is $10'\times 6.5'$, similar in radio and X-ray, elongated in the east-west direction. At its often-quoted distance of $3.2$~kpc \citep{Roberts_Goss+93a} it has a size of $\sim 9\times 6$~pc; a more recent analysis estimates $d \sim 2$~kpc \citep{Kothes13a}. This supernova remnant has been traditionally associated  with SN 1181 \citep{Stephenson_Green02a}. However, the expansion of the optical filaments surrounding the source \citep{Bietenholz06a,Fesen_Rudie+08a}, casts doubt on this association \citep{Chevalier05a}, and the 1181 event is now linked with the newly discovered remnant Pa 30 \citep{2021ApJ...918L..33R,Schaefer23a}. The age of 3C~58 is thus uncertain.\\
\\
In radio the morphology of the nebula is quite complex with many loops and filaments, with typical width $\sim 2$\,arcseconds and luminosity contrast of $\sim 2\times$ \citep{Reynolds_Aller88a,Bietenholz06a,Castelletti19a}. A prominent filament is found close to the pulsar, but unlike the Crab this does not appear to mark the  pulsar wind termination shock \citep{Bietenholz06a}. The radio spectral index of $-0.05$ is quite uniform over the nebula  \citep{Bietenholz_Kassim+01a}. The non-thermal radio nebula is embedded in a network of thermally emitting and radially expanding knots \citep{Bocchino_Warwick+01a,Gotthelf_Helfand+07a,Fesen_Rudie+08a} hosting an estimated total mass $\sim 0.1$~M$_\odot$.\\
\\
In X-rays 3C 58 displays a central jet–torus structure \citep{Slane_Helfand+02a}, clear evidence of energy injection from its central pulsar PSR J0205+6449. The torus has been detected in the IR \citep{Slane_Helfand+08a}. An optical detection is also claimed \citep{Shibanov_Lundqvist+08a}; this is slightly fainter and softer than predicted from a simple extrapolation of the X-ray spectrum. The torus is oriented north-south, suggesting that the spin axis of the pulsar is aligned east-west and, unlike other PWNe, is viewed almost edge-on ($\pm 15^\circ$) \citep[e.g.][]{Ng_Romani04a}. The total nebula [2-8]~keV luminosity is $\sim 9\times 10^{-12}$~erg~s$^{-1}$~cm$^{-2}$ \footnote{See {\sl Chandra} Supernova Remnant Catalog entry at \url{https://hea-www.harvard.edu/ChamdraSNR/G130.7+03.1/}}; about 20\% comes from the torus. The spectrum softens with radius, with photon index $\Gamma$ increasing from $\sim2$ to $\sim$2.5 \citep{An19a}. Spectral modeling of the source suggests a magnetic field in the range $30-40$~$\mu$G \citep{Bucciantini_Arons+11a,Torres_Cillis+13a,Lu_Gao+17a,Kim_Park+19a}. The body of the PWN shows an X-ray morphology characterized by filaments or loops, closely following similar radio features. \\
\\
Radio polarization maps \citep{Wilson_Weiler76a,Reich02a,Sun_Reich+11a,Castelletti19a} show patchy polarization, with a typical coherent scale of about 10\% of the nebula. The inferred magnetic field orientation follows the filamentary structure, with a global tendency toward alignment with the nebula major axis. The average polarization degree (PD) in the radio is $\sim 6-15$\%, with some regions reaching PD$_{\rm max} \approx 30$\%, and others showing negligible polarization.  The central zone of the radio nebula, coincident with the X-ray torus shows radio PD up to a local maximum of $ \sim 10-13$\%, with an electric vector polarization angle (EVPA) in the east-west direction. \\
\\
PSR J0205+6449 has also been detected at radio, X-ray and GeV energies \citep{Camilo_Stairs+02a,Murray_Slane+02a,Livingstone_Ransom+09a,Kuiper_Hermsen+10a,Abdo_Ackermann+09c,Li+2018A} and there is a tentative optical counterpart \citep{Shearer_Neustroev08a,Moran_Mignani+13a}. Its period $P = 65.68$~ms, and  period derivative $\dot{P} = 1.94\times 10^{-13}$~s~s$^{-1}$, correspond to a spin-down power $\dot{E} \sim 2.7\times 10^{37}$~erg~s$^{-1}$, a spin-down characteristic age $\tau \sim 5400$~yr, and an effective surface dipole magnetic field $B_{\rm s} = 3.6\times 10^{12}$~G. Its Dispersion Measure $DM=140.7\,{\rm cm^{-3}pc}$ corresponds to $d \sim 2.8$\,kpc 
%implies a distance of $\sim 4.5$~kpc, according to the standard electron density distribution \citep{Cordes_Lazio02a}, larger than that inferred for the nebula based on HI absorption \citep{Roberts_Goss+93a,Kothes13a}. However more recent models 
\citep{Yao+17a}, consistent with the more modern SNR distance estimate above.
%; in this paper we adopt the 2\,kpc \citep{Kothes13a} values
The pulse profile in X-rays is characterized by a very narrow (FWHM$_\phi \sim 0.02$) main peak that lags the radio pulse by $\delta_\phi \sim 0.09$. There is evidence for a $\sim 5\times$ weaker, but $\sim 2\times$ broader, secondary X-ray peak lagging the primary by $\Delta_\phi \approx 0.5$. The peak flux ratio increases toward higher energies with the secondary peak dominating above 1~GeV. The pulsed X-ray emission can be fitted with a single power-law with photon index $\Gamma \sim 1.0$ and integrated pulsed flux $f_{2-8keV} \sim 2.3 \times 10^{-13}$~erg~s$^{-1}$~cm$^{-2}$. \\
\\
Sect.~\ref{sec:obs} presents the polarization observations, both X-ray and radio, used in this work. An analysis of the unusually strong X-ray polrization background is given in Sect.~\ref{sec:bkg}. Sect.~\ref{sec:analisys} reports our findings for the PSR and PWN polarization, and in Sect.~\ref{sec:concl} we present our conclusions.

\section{Observations}
\label{sec:obs}
The Imaging X-ray Polarimetry Explorer (IXPE)%, a groundbreaking space observatory, 
was successfully launched on December 9, 2021 \citep{Weisskopf_Soffitta+22a}. As the first mission solely dedicated to imaging X-ray polarization, IXPE is a collaborative effort between NASA and the Italian Space Agency (ASI). The observatory comprises three identical telescopes, each outfitted with a polarization-sensitive Gas Pixel Detector (GPD) positioned at its focal plane. These GPDs are housed within a detector unit (DU).\\
\\
\src was observed by IXPE twice in the summer of 2024.  The first segment began on July 9, 2024, at 21:54:30 UTC and lasted until July 14, 2024, at 14:53:13 UTC. The second segment started on July 29, 2024, at 04:13:54 UTC and continued until August 10, 2024, at  04:42:04 UTC. These segments were then combined to create a single data set with approximately 966 ks \texttt{LIVETIME}. IXPE Level-2 data are publicly available for download on the HEASARC archive\footnote{\url{https://heasarc.gsfc.nasa.gov/docs/ixpe/archive/}} (ObsID 03002099). Due to the high level of solar activity during both segments, data were reprocessed with a non-standard approach to reduce contamination by flares (see Sect.~\ref{sec:bkg}). The arrival times of the remaining events were corrected to the solar system barycenter using the \texttt{HEASOFT BARYCORR} tool and the \texttt{JPL 421} ephemeris. The absolute sky coordinates were bore-sighted by adjusting the FITS WCS keywords in order to align the IXPE brightness peak, with the intensity peak of {\sl Chandra} images. \\
\\
We also analyzed a 2.69 hour 2017 \src JVLA radio observation in the S ($\nu_c =3$\,GHz, $\Delta \nu=2$\,GHz, i.e.\,$\lambda\approx$6~cm) and C ($\nu_c =6$\,GHz, $\Delta \nu=4$\, i.e.\,$\lambda\approx$3~cm) bands. The S band data included configurations C and D, and the C band configuration C. The data were taken in continuum mode with all four Stokes parameters. Combining all data, the minimum baseline is 35\,m, so that the 6 and 3~cm observations are sensitive to angular scales up to 490\arcsec\ and 240\arcsec, respectively. J0137+3309 was observed for bandpass, polarization angle, and flux density calibrations; J0217+7333 was observed as a calibrator for the visibility phases; J0319+4130 was observed to monitor the polarization leakage. \\
\\
All JVLA data analysis used the Common Astronomical Software Application (\texttt{CASA}) package \citep{CASA22a}\footnote{\url{https://casa.nrao.edu/}}. We first applied the JVLA CASA Calibration pipeline (6.6.1-17) to remove radio frequency interference (RFI), did basic flux, bandpass, gain, and polarization calibrations and determined the delay. We solved for the multiband cross-hand delays of the two basebands separately. We then calculated the polarization leakage terms and the R-L polarization angle.  Additional RFI flagging was performed after the pipeline processing.  After the flagging and calibration process, we formed Stokes I, Q, and U images, deconvolved using the task \texttt{tclean}. To compromise between sensitivity and resolution, we used the Briggs weighting algorithm with \texttt{robust} $= 0.5$ for these two-band images. After the initial round of cleaning, we interactively applied a clean mask, to restrict the cleaned area in subsequent iterations. The synthesized beams in the final map are $8.3''\times 6.3''$ in S band and $3.49''\times 3.01''$ in C band. The rms noise levels in the Stokes I images are 83\,mJy\,beam$^{-1}$ (S) and 72\,mJy\,beam$^{-1}$ (C). The noise levels in Stokes Q and U are 80\,mJy and 50\,mJy, respectively, similar to the theoretical rms. \\  
\\
The rotation measure (RM) of these 2 bands is obtained by splitting the data into 4 sub-band images, then using the task \texttt{RMFIT} to determine the RM map. We correct the Q and U Stokes maps to infinite frequency using this RM map.

\section{Background}
\label{sec:bkg}
\begin{figure}
   \centering
   \includegraphics[width=8cm]{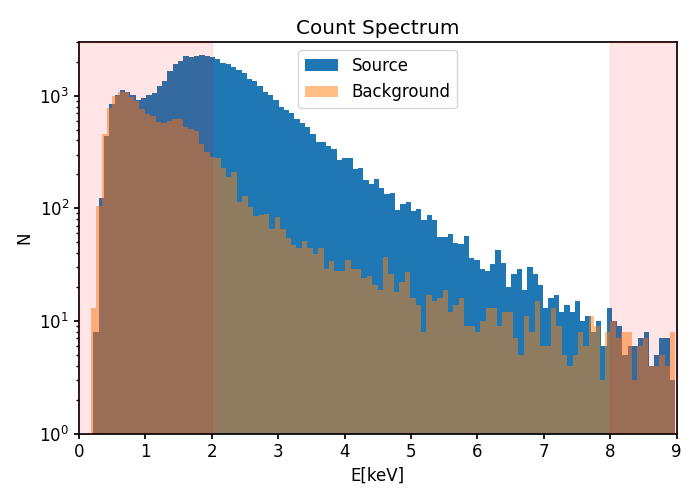}
   \caption{Comparison of the count spectrum over a region of $50$~arcsec radius centered on the source vs an equal size background region, for the background rejected data before de-flaring. The red shaded regions highlight counts outside the fiducial [2-8]~keV energy range of IXPE calibration.}
    \label{fig:src_vs_bkg.png}    
    \end{figure}

The \src observation period was characterized by unusually strong solar flaring activity. X-rays from solar flares can contribute a polarized signal to the background. Initial evidence for possible solar flare contamination  was seen in the level of the background count rate in Detector Unit 2 (DU2), which was found to be 20\% higher than in Detector Unit 1 (DU1), which was partly shielded from the Sun during this observation.\\  
\\
For background event rejection we adopted a strategy following the approach introduced in \cite{Di_Marco+23a}. However, for additional background suppression we modified their (highly conservative) rejection criteria to a more aggressive recipe, tuned specifically for \src:
\begin{align}
0.004 E +0.88 [1-e^{-(E+0.25)/1.1)}]+0.01 < EVT\_FRA <1\\
NUM\_PIX < 115 + 195  \times (E/8)^2
\end{align}
where $E$ is the energy of the event in keV, and $EVT\_FRA$ and $NUM\_PIX$ are parameters characterizing the track in Level-1 files. This allowed us to remove $\sim 10$\% more events than by applying standard cuts.\\
%Maybe should give the apparent % of lost source counts? RWR
\\
Comparing the count spectrum in a region of 50~arcsec radius centered on the pulsar position, with that from a same-sized background region, one can see that above $\sim$7~keV the background dominates (Fig.~\ref{fig:src_vs_bkg.png}). Thus we decided to limit further analysis to the [2-6]~keV energy range.\\
%%%%%%%%%%%%%%%%%%%%%%%%%%%%%%%%%%%%%%%%%%%%%
%%%%%%%%%%%%%%%%%%%%%%%%%%%%%%%%%%%%%%%%%%%%%
\\
In order to assess residual contamination by flaring activity, we performed a polarization analysis of the background. We selected all events in the Field of View (FoV) $>120^{\prime\prime}$ from the PSR, and used the \texttt{ixpeobssim} (31.0.2) \texttt{PCUBE} algorithm to measure the background polarization. By simulating this system using the \texttt{ixpeobssim} package, and archival {\sl Chandra} images, we verified that the PWN contributes less than 2\% of the total events beyond a distance of $120^{\prime\prime}$ from the PSR. Combining the three DUs, this background was indeed significantly polarized, with polarization degree PD$=11.1\% \pm 1.3\%$ at polarization angle of $-11.0^\circ \pm 3.3^\circ$. However, the signal varies greatly between the DUs, a clear indication of solar contamination (see Tab.~\ref{tab:bkg}). Similar results (but at lower statistical significance) were obtained for smaller background regions, further from the detector edges, so this is not a vignetting effect.\\

To confirm that the signal is source-independent, we used the IXPE's Science Operation Center reconstruction pipeline\footnote{\href{https://heasarc.gsfc.nasa.gov/docs/ixpe/analysis/IXPE-SOC-DOC-009-UserGuide-Software.pdf}{https://heasarc.gsfc.nasa.gov/docs/ixpe/analysis/IXPE-SOC-DOC-009-UserGuide-Software.pdf}} and the \texttt{STATUS2} flags to select data from occultation periods when no calibration source was being exposed, giving a Level-2 format data set that could be exploited as a pure background sample. As it happened, flaring activity during these background windows was especially strong, leading to a higher polarization degree (Silvestri, private communication). The same analysis run on other archival sources reveals that a significant modulation of the background is often recovered during Solar flaring activity, both in and out of occultation, with similar polarization position angles.   \\
\\
Thus standard background rejection is insufficient, even with our more aggressive cuts. For this reason we de-flared our observations by removing all time intervals when the count rate of the background was 2.5 times higher than the median rate. This excised all events recorded during class M solar flares, as measured by the Geostationary Operational Environmental Satellites (\textit{GOES})  \citep{Hanser_Sellers+96a}\footnote{Archival data for solar flaring activity of the GOES satellites can be found at \url{https://www.spaceweatherlive.com/en/solar-activity/solar-flares.html}}. The excised time amounts to about 5\% of the total observation. We repeated the analysis on this de-flared dataset and found a polarization degree of $6.8\% \pm 1.4\%$ and a polarization angle of $-6.1^\circ \pm 5.9^\circ$, still highly significant. However, lowering the flare excision threshold to twice the median rate did not substantially decrease the background polarization (still PD$=5.9\% \pm 1.4\%$). This indicates that the flare contamination was caused by continuous low level activity covering the entire first segment of the observation. Unfortunately this activity was too steady for count rate cuts to allow effective excision. 
\begin{table}
     $$ 
         \begin{array}{p{0.2\linewidth}l c c r}
            \hline
            \noalign{\smallskip}
            Unit      &  PD[10^{-2}] & PA & MDP\_99 \\
            \noalign{\smallskip}
            \hline
            \noalign{\smallskip}
            DU1\_rj & 3.7\pm 2.4 & 40.2^\circ \pm 18.7^\circ & 0.073  \\
            DU2\_rj & 16.9\pm 2.1 & -14.8^\circ \pm 3.6^\circ & 0.065 \\
            DU3\_rj & 14.5\pm 2.2 & -11.8^\circ \pm 4.3^\circ & 0.067 \\
            TOT\_rj & 11.1\pm 1.3 & -11.0^\circ \pm 3.3^\circ & 0.039 \\
            DU1\_df & 4.1\pm 2.5 & 48.0^\circ \pm 17.5^\circ & 0.076  \\
            DU2\_df & 10.0\pm 2.4 & -11.0^\circ \pm 6.9^\circ & 0.072  \\
            DU3\_df & 10.8 \pm 2.4 & -10.4^\circ \pm 6.3^\circ & 0.072   \\
            TOT\_df& 6.8\pm 1.4 & -6.1^\circ \pm 5.9^\circ & 0.042 \\
            \noalign{\smallskip}
            \hline
         \end{array}
     $$ 
           \caption[]{Background Polarization properties derived over the entire FoV at a distance > 120 arcsec from the PSR. Results for each of the three IXPE Detector Units (DUs), and for the total combined background (TOT), both after rejection (rj) and after deflaring (df). Errors are at 1$\sigma$. $MDP\_99$ is the minimum detectable polarization at 99\% confidence level.  }
         \label{tab:bkg}
\end{table}

\begin{figure*}
    \centering
    \hspace{-1em}
    \includegraphics[width=0.33\linewidth, bb=-40 0 420 430, clip]{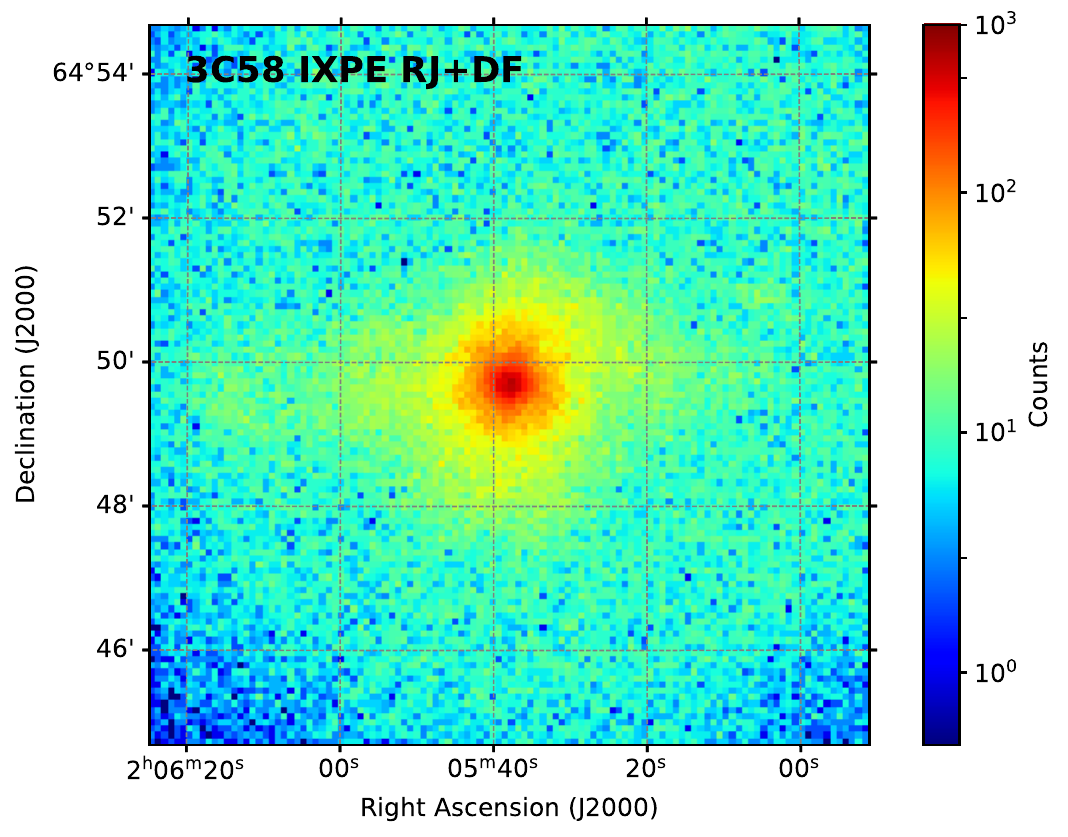}\includegraphics[width=0.33\linewidth, bb=60 0 520 430, clip]{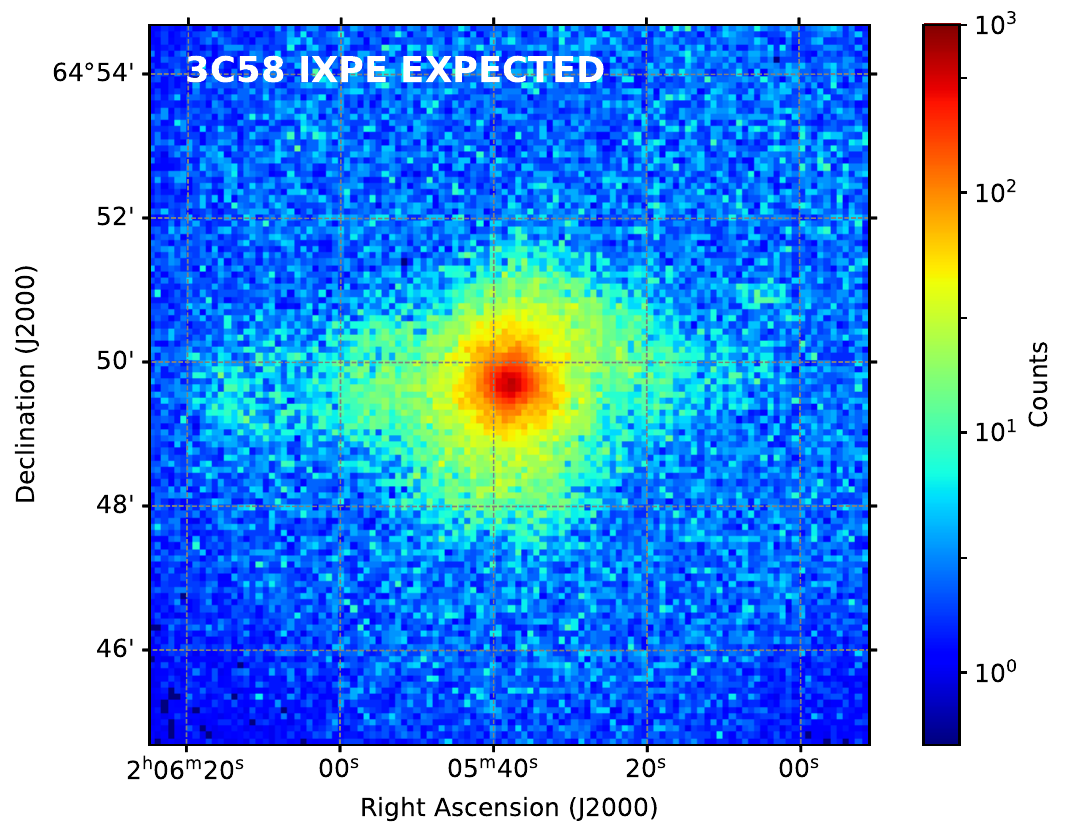}\includegraphics[width=0.33\linewidth,bb=60 0 520 430, clip]{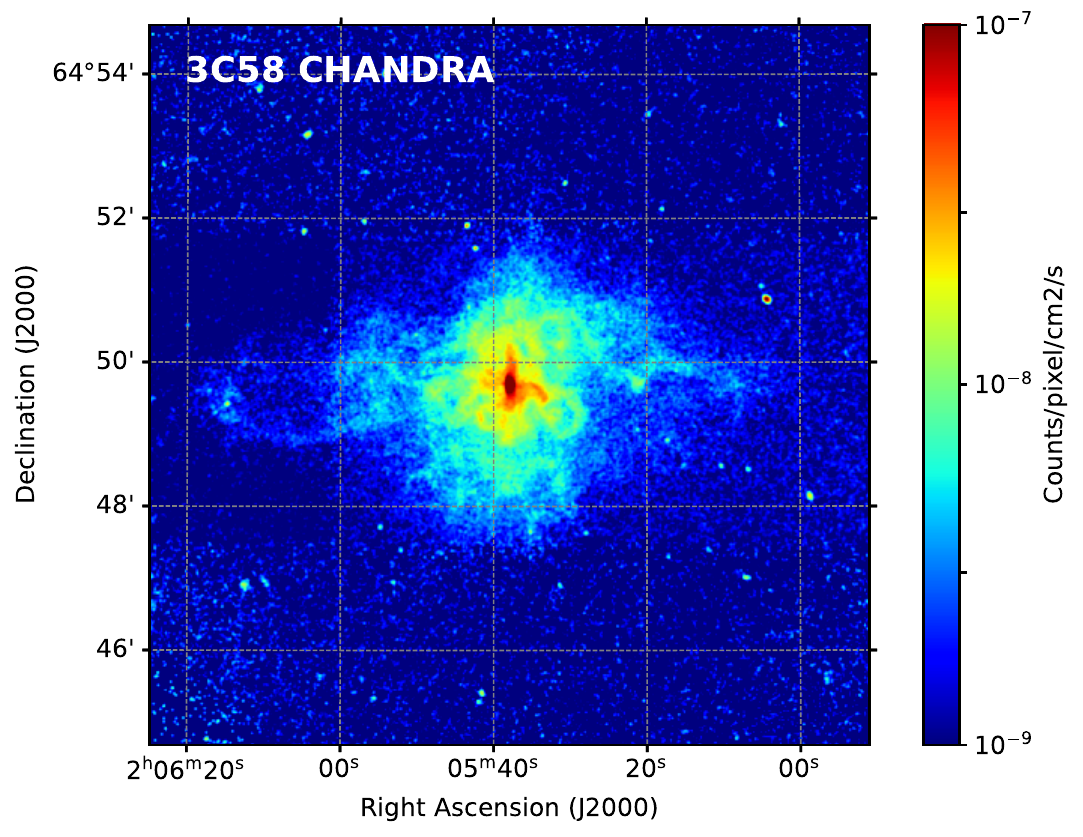}
    \caption{Left panel: IXPE total count map in the full [2-8]~keV band after background rejection and de-flaring. Central panel: maximum likelihood expectation value for the source IXPE counts in the [2-8]~keV range, using the method of \cite{2022MNRAS.515.5185E}. Right panel: {\sl Chandra} count flux in the in the [2-8]~keV band derived from ObsIds: 728, 3832, 4382, 4383, 25786, 25787, 25788, 25789, 25790, 26402.  }
    \label{fig:counts}
\end{figure*}

\section{Analysis}
\label{sec:analisys}
\subsection{Imaging and Polarization}
\label{sec:imaging}
\begin{figure}
    \centering
    \includegraphics[width=0.99\linewidth]{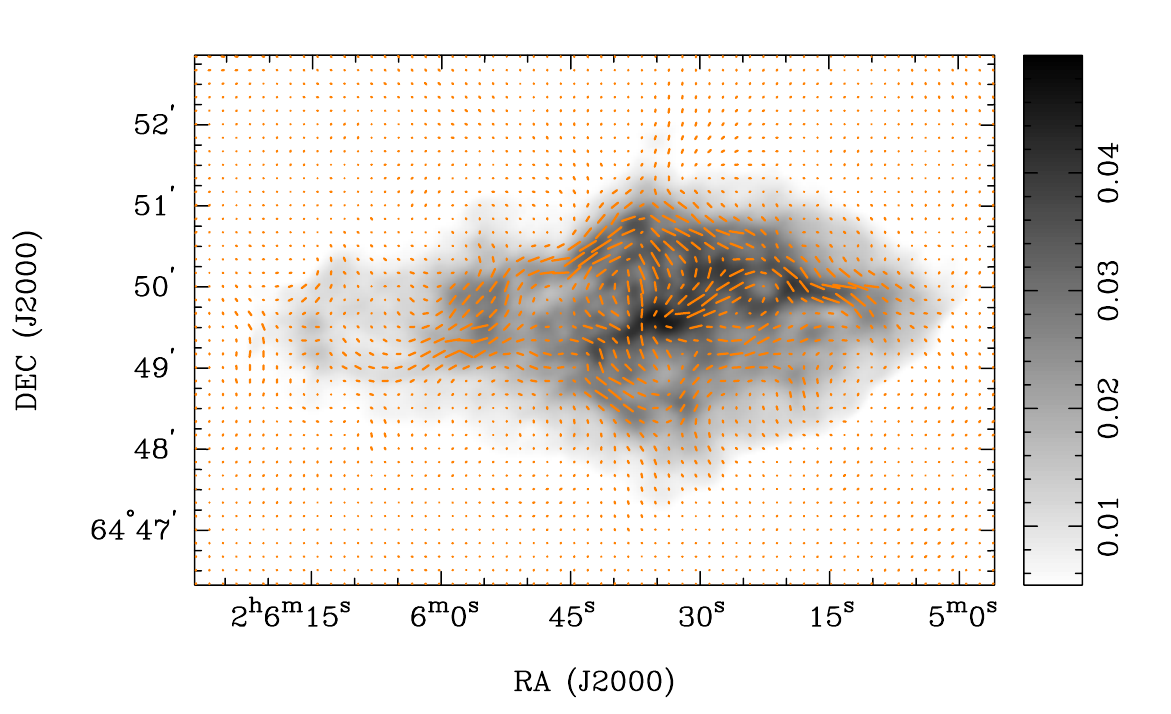}
    \includegraphics[width=0.99\linewidth]{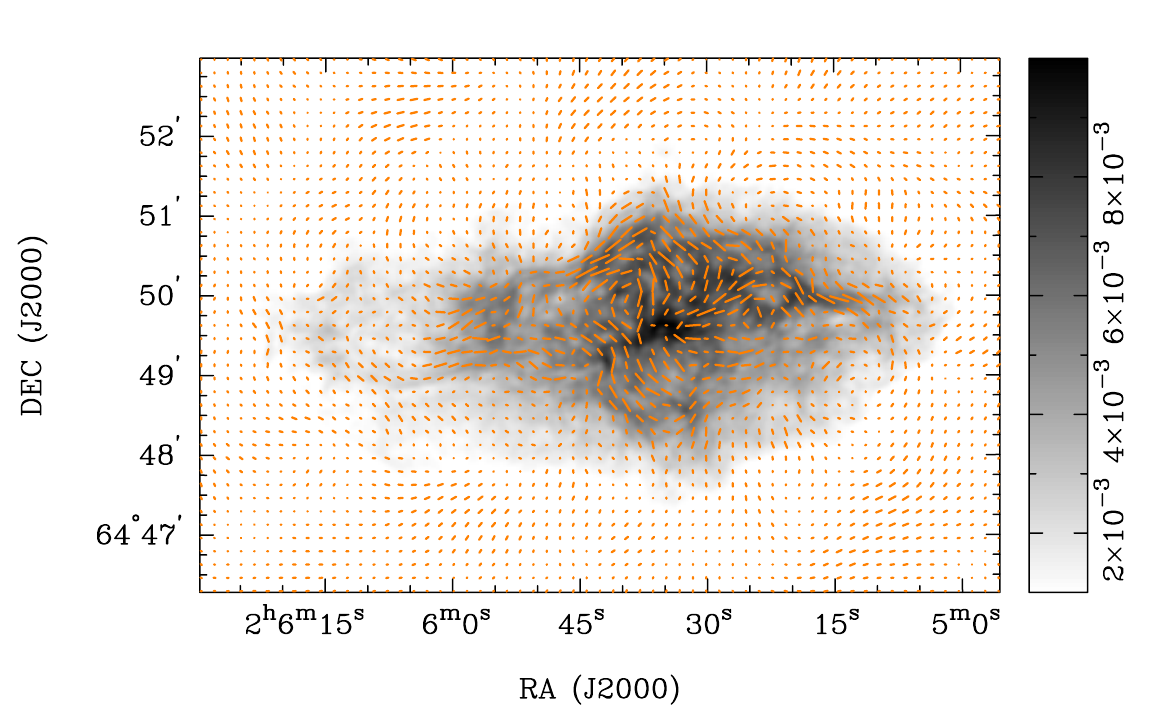}
    \caption{Upper panel: VLA S-band polarization map of \src. Magnetic field direction overlayed on the total intensity Jy per beam. Lower panel:the same for the C-band.}
    \label{fig:radio}
\end{figure}

\begin{figure}
\centering
    \includegraphics[width=0.8\linewidth]{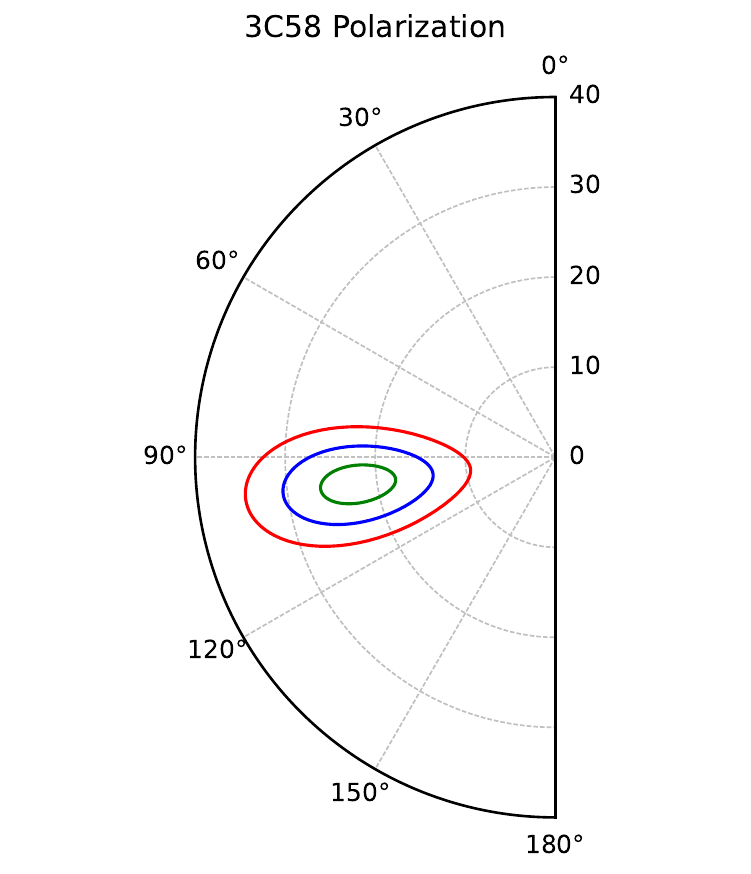}
    \includegraphics[width=0.8\linewidth]{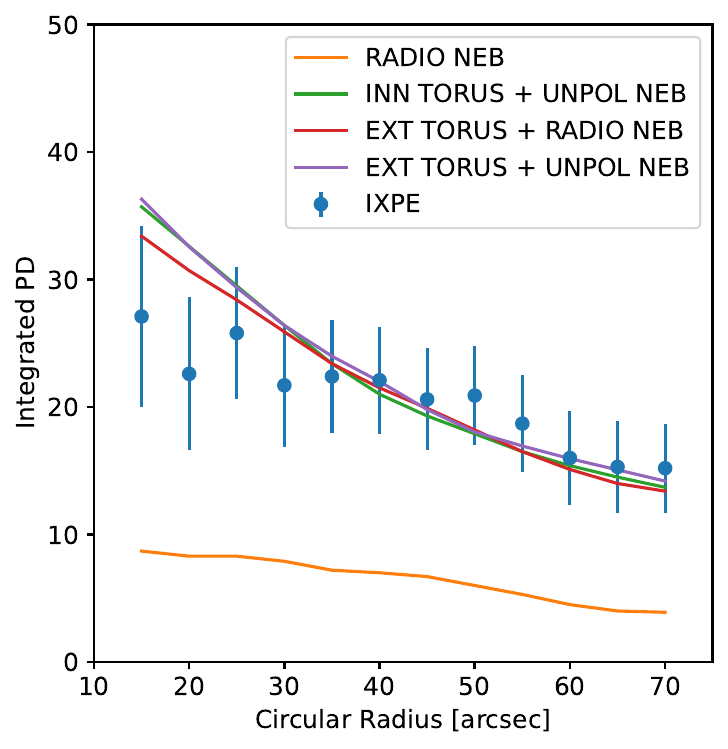}
    \caption{Upper panel: polarization angle and degree in a region of 40~arcsec radius centered on the pulsar in the [2-6]~keV energy range. Contours represent the 1,2 and 3$\sigma$ confidence regions. Lower panel: polarization degree as a function of the region size (for regions centered on the PSR). Points represent IXPE values with their 1$\sigma$ uncertainty. Solid lines represent the expectation for various models: a polarized inner torus (PD=75\%) embedded into an unpolarized external nebula (green); a polarized extended torus (PD=55\%) embedded into an unpolarized external nebula (purple); a polarized extended torus (PD=50\%) embedded into an external nebula with the same polarization as seen in radio (red); a nebula having the same polarization as seen in radio (orange).}
    \label{fig:pol}
\end{figure}

In Fig.~\ref{fig:counts} the total count map of the target region after background rejection and de-flaring is shown. In comparison with {\sl Chandra} images, it is evident that IXPE cannot resolve either the inner jet-torus structure, or any of the PWN filament/loop structures. However, the large scale PWN morphology is well captured. In particular, the maximum likelihood expectation map \citep{2022MNRAS.515.5185E} for the source emission clearly shows that the large-scale features are indeed observed. Still, the counts in the outer regions of the nebula are too few to attempt a detailed morphological polarization study, especially since bright unresolved sources, like the PWN core, produce spurious polarized halos due to polarization leakage \citep{Bucciantini23b}. Note also that the source covers a large fraction of the FoV, with the eastern lobe's outer edge affected by vignetting.\\
\\
In Fig.~\ref{fig:radio} we show the radio polarization map of \src in the JVLA 2-4~GHz S band, and 4-8~GHz C band. The radio polarization structure is similar at the two frequencies and the magnetic field follows the structure of the radio loops in the body of the nebula. Note that the radio polarization does not show the central torus or jet. The polarization degree in the central region corresponding to those bright X-ray features is $\approx 10\%$, with a polarization angle of $\sim 95^\circ$, aligned with the (E-W) PWN main axis. This is compatible with a toroidal magnetic field in the central region. Regions of higher polarization up to $\sim 30\%$ are found only at the edges of the PWN. \\
%%%%%%%%%%%%%%%%%%%%%%%%%%%%%%%%%%%%%%%%%%%%%
%%%%%%%%%%%%%%%%%%%%%%%%%%%%%%%%%%%%%%%%%%%%%
\\
The {\sl Chandra} map shows that the X-ray emission of \src is dominated by the central torus region, whose extent is much smaller than the {\it IXPE} ($\sim30^{\prime\prime}$ Half Power Diameter) resolution. In order to evaluate the possible spatial dependence of the X-ray polarization in \src, and its compatibility with a point-like source, a polarization analysis using the \texttt{PCUBE} algorithm of \texttt{ixpeobssim} was performed over circular regions centered on the PSR position with radius varying from 15 to 60~arcsec. The background-subtracted polarization degree (PD) was found to have a decreasing trend with aperture size from 27\% at 15~arcsec to 16\% at 60~arcsec, with a significance ranging from 4.0$\sigma$ to 5.3$\sigma$ (the maximum is achieved for a region of 40~arcsec radius), while the polarization angle (PA) is always $\sim 100^\circ$. Results are reported in Tab~\ref{tab:table2} for a circular region of 40~arcsec radius centered on the PSR and in Fig.~\ref{fig:pol} as a function of the region size. \\
\\
%%%%%%%%%%%%%%%%%%%%%%%%%%%%%%%%%%%%%%%%%%%%%%%%%%%%%%%%%%%%%
%%%%%%%%%%%%%%%%%%%%%%%%%%%%%%%%%%%%%%%%%%%%%%%%%%%%%%%%%%%%%
The analysis was repeated running a spectro-polarimetric fit with both \texttt{3ML}~\citep{2015arXiv150708343V} and \texttt{XSPEC}, with an independent background selection. Both tools implement a forward folding technique, which allows one to fit the Stokes I, Q and U spectra, and retrieve the best-fit spectral and polarimetric models. \\
\\
The spectro-polarimetric \texttt{XSPEC} analysis was performed using the HEASoft software package (version 6.34)\footnote{\url{https://heasarc.gsfc.nasa.gov/docs/software/heasoft/}}. The source ($r < 40$~arcsec from the PSR) and background (a circular region of $120$~arcsec radius centered at RA=2:05:36, DEC=64:53:31) regions were filtered with the \texttt{FTOOLS} command \texttt{xselect}, setting the parameter \texttt{Stokes = NEFF} to extract the weighted Stokes I, Q and U spectra for all three detector units. Response files for the I, Q, and U spectra were generated using the IXPE-specific command \texttt{ixpecalcarf}, which utilized the cleaned Level-2 event lists for each DU and the corresponding attitude files from housekeeping data.
Data fitting was performed using \texttt{XSPEC} (version 12.14.1)\footnote{\url{https://heasarc.gsfc.nasa.gov/xanadu/xspec/}}. The Stokes I, Q, and U spectra (background-subtracted) were simultaneously fitted across all three DUs using a model comprising absorption (the \texttt{tbabs} by \citet{Wilms+20a}), a power-law (\texttt{powerlaw}), and constant polarization degree and angle (\texttt{polconst}). To account for flux calibration differences between DUs, a cross-normalization constant (\texttt{const}) was included in the model. For the line-of-sight absorption toward the source, we fixed the hydrogen column density \(N_\mathrm{H}\) at \(0.42 \times 10^{22} \, \mathrm{cm}^{-2}\) \citep{Picquenot+24a}. Within our assumed 2--6 keV range, an absorbed power-law with a photon index of $2.25 \pm 0.03$ adequately describes the spectrum. The unabsorbed flux extrapolated in the 2--8 keV range is estimated to be approximately \((2.82 \pm 0.05) \times 10^{-12} \, \mathrm{erg \, cm^{-2} \, s^{-1}}\) at 1$\sigma$ confidence level. The polarization degree was determined to be $21.4\% \pm 3.5\%$, while the polarization angle was found to be $98.1^\circ \pm 4.8^\circ$). A summary of the best-fit parameters is provided in Tab.~\ref{tab:table2}. \\ 
\\
For the \texttt{3ML} analysis the Stokes spectral files for the source ($r\,<\,40^{\prime\prime}$ from the PSR) and background ($120\,<\,r\,<\,300^{\prime\prime}$ from the PSR) regions were created with the \texttt{PHA1}, \texttt{PHA1Q} and \texttt{PHA1U} algorithms of \texttt{ixpeobssim} using an unweighted binning scheme. For the response files, we used the on-axis response (version 13 and validity date $20240701$) provided in the CALDB folder of \texttt{ixpeobssim}. As in the \texttt{XSPEC} analysis the spectral model was an absorbed power-law, with absorption given by the \texttt{PhAbs} model with $N_H=0.42 \times 10^{22} \; \mathrm{cm}^{-2}$. The polarization degree and angle of the source were modeled as constant in the 2--6~keV energy range. A multiplicative normalization was included to cross-calibrate between the different DUs. The resulting best-fit photon index was found to be $2.31 \pm 0.03$ with a polarization of $22\% \pm 4\%$ at an angle of $98^\circ \pm 5^\circ$, in good agreement with the previous analysis. These polarization results are summarized in Tab.~\ref{tab:table2}. \\

\begin{table*}

\end{table*}

\begin{table*}
     $$ 
         \begin{array}{p{0.2\linewidth}l c c c}
            \hline
            \noalign{\smallskip}
            Method      &  Q/I & U/I & PD[10^{-2}] & PA  \\
            \hline
            \noalign{\smallskip}
            \texttt{ixpeobssim} & -0.213\pm 0.042 & -0.059 \pm 0.042 & 22.1 \pm 4.2 & 97.7^\circ \pm 5.5^\circ\\
            \texttt{XSPEC} & -0.204\pm 0.035 & -0.059 \pm 0.035 & 21.4\pm 3.5 & 98.1^\circ \pm 4.8^\circ\\
            \texttt{3ML} & -0.21 \pm 0.04  & -0.06 \pm 0.04 & 22.0 \pm 4.0 & 98.0^\circ \pm 5.0^\circ\\
            Simultaneous Fitting & -0.171 \pm 0.033 & -0.054 \pm 0.033 & 17.9 \pm 3.3 & 98.7^\circ \pm 5.2\\
         \end{array}
     $$
           \caption[]{Background subtracted polarization properties of the inner 40~arcsec region of 3C 58 (except for simultaneous fitting, where the extraction region is $75" \times 75"$ and the pulsar polarization is fitted simultaneously with the nebula polarization). Errors are at 1$\sigma$ level. For the \texttt{XSPEC} analysis the reduced chi-square value is $\chi^2/\text{d.o.f.} = 869.48/885$. For \texttt{3ML} analysis the reduced chi-square value is $\chi^2/\text{d.o.f.} = 916.72/885$ ($p$-value = 0.22). For simulatneous fitting, the reduced chi-square value is $\chi^2/\text{d.o.f.} = 522 / 450$ ($p$-value = 0.01)}.
           \label{tab:table2}
\end{table*}     

\subsection{Timing} 
\label{sec:timing}
\begin{figure}
    \centering
    \includegraphics[width=0.9\linewidth]{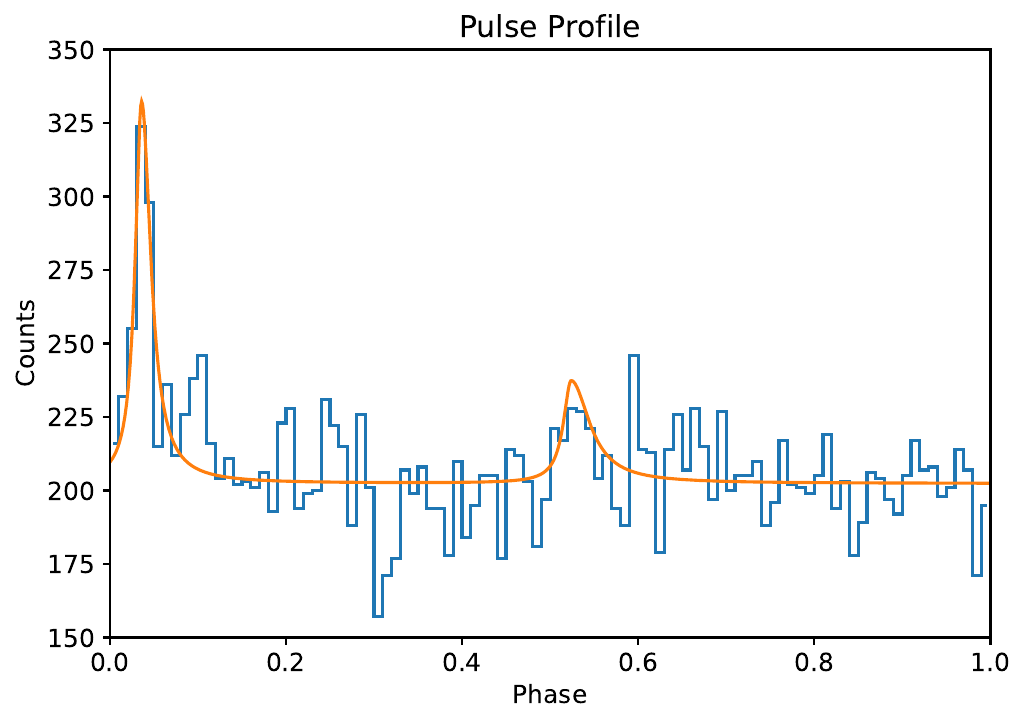}\\
    \includegraphics[width=0.9\linewidth]{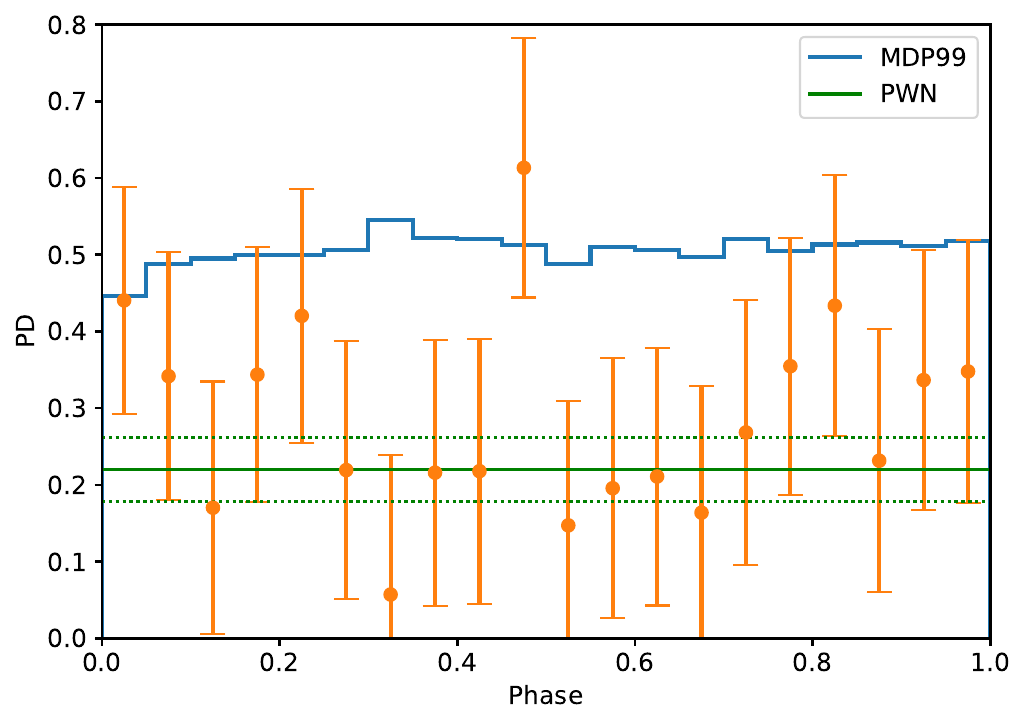}
    \caption{Upper panel: IXPE pulse profile for the PSR J0205+6449 / \src system, in the [2-6]~keV band within a region of 40~arcsec radius from the PSR, compared to the analytical model by \citet{Kuiper_Hermsen+10a}. Lower panel: phase resolved polarization degree within the same region, together with the minimum detectable polarization at 99\% confidence level. }
    \label{fig:pp}
\end{figure}

For a phase analysis we needed a current pulsar ephemeris. Given that the ephemeris of the Australia Telescope National Facility (ATNF) catalog\footnote{accessible at \url{https://www.atnf.csiro.au/research/pulsar/psrcat/}} is 17 years old, the ephemeris for the current observation was measured from {\it Fermi}-LAT $\gamma$-ray data. \src is bright in the gamma-ray band so we used the 100\,MeV--100\,GeV LAT data with a 1-degree region of interest to meausre the pulsations. Times of arrival were accumulated in 10 day exposure intervals, and a phase-coherent timing analysis is performed using \texttt{Tempo2} \citep{Hobbs+24a} provided an up-to-date ephemeris, reported in Tab~\ref{tab:ephem}. The ephemeris used in this paper covers the time range from MJD 60420 to 60550 (corresponding to April 20, 2024 - August 28, 2024). No evidence for any glitch was found during this period. In Fig.~\ref{fig:pp} we show the phase folded {\it IXPE} pulse profile in the [2-6]~keV energy band over a region of 40~arcsec radius centered on the PSR, together with the analytical model of \citet{Kuiper_Hermsen+10a}. The main peak is clearly evident, although the second peak is not well detected. In order to assess possible pulsar polarization, we performed a phase-resolved polarization analysis dividing the period into 20 equally spaced phase bins and applying to each the \texttt{PCUBE} method of \texttt{ixpeobssim}. In all phase bins the PD is below the $MDP_{99}$, and consistent within $\sim 1\sigma$ with the PD value inferred for the PWN. No significant increase or drop in PD is seen in correspondence with the main peak. A value of PD higher than the $MDP_{99}$ is only found in one bin just ahead of the second pulse. However, once counting statistics is taken care of, the chance of having one out of 20 bins above $MDP_{99}$ is found to be $\sim 18$\%. Indeed such high polarization bin disappears if a slightly different binning is used. We also attempted to measure the polarization properties in phase bins matching both the main and second peak, but found no significant difference with respect to the off-pulse values, in terms of polarization degree and angle. \\
\\

\begin{table}
\[
         \begin{array}{p{0.5\linewidth}l r }
            \hline
            \noalign{\smallskip}
            PEPOCH [MJD] &  60550 \\
            F0 ~{\rm [Hz]}     &  15.19218577(14)\pm (20) \\
            F1  &  -4.483(82)\pm (12)\;\;\;\times 10^{-11}\\
            F2 ~{\rm [1/Hz]} & 1.(89)\pm(78)\;\;\;\times 10^{-21}   \\
            \noalign{\smallskip}
            \hline
         \end{array}
         \]
           \caption[]{PSR ephemeris determined for the $\gamma$-ray times of arrival. In brackets the last significative digits and their errors as determined by \texttt{Tempo2} }
         \label{tab:ephem}
\end{table}

\subsection{A Search for Finer Polarization Structure} 

With the pulse ephemeris above we attempted to search for finer polarization effects, using the \textit{``simultaneous fitting''} procedure described by \cite{Wong2023}. Unlike the simple on-off method, the method utilizes externally ({\it Chandra})-constrained models of the PSR and PWN fluxes to help simultaneously solve for their polarization. These models are generated using a high spatial-resolution {\sl Chandra}-derived template for the nebular flux, and the pulsar phase-varying X-ray spectrum. It convolves these with the IXPE instrument response to build the predicted Stokes I map for each component in various pulsar phase bins. The flux models and IXPE polarization data are binned in phase, energy, and spatial coordinates to give an overdetermined linear set of equations, in which the unknown parameters are the polarization of the pulsar, which only varies with phase, and the nebula, which only varies spatially. These can be solved analytically via least-squares regression. For this analysis, we compiled 10 {\sl Chandra} archival observations (ObsIDs 728, 2604, 3832, 4382, 4383, 25786, 25787, 25788, 25789, 25790) with aggregate {$\sim$\,500} ks exposure (see Fig. \ref{fig:counts} for the merged {\sl Chandra} image) and replaced the pulsar-dominated peak ($\rm r=0.5^{\prime\prime}$, $2\times$FWHM) with a sub-sampling of nebular events in a $1-1.5^{\prime\prime}$ annulus. This is our high-resolution nebula model. The pulsar light curve was obtained by using the analytic model of \cite{Kuiper_Hermsen+10a} (their Eq. 2 \& 3 and Table 3), and the spectrum of the two pulses were described using the PCA measurements reported in Tab. 4, with the normalization converted to the nominal IXPE $2-8$\,keV energy range. We designated any flux below 1\% of the maximum flux of each pulse as belonging to a constant `off-pulse'' component, with photon index $\Gamma = 2.42$ as described by \cite{Kim_An21a} (see Tab. 2). These fluxes were convolved with the {\it IXPE} effective area and the on-axis PSFs derived by \cite{Dinsmore24} in \texttt{ixpeobssim} to obtain the predicted IXPE counts for each component.  \\
\\
A few correction factors were needed to match the simulated and observed light curves and off-pulse images. First, we rescaled the nebular component by $0.8\times$ to match the DC levels of the total light curve. Then we scaled the pulsar component by 0.68$\times$ to match the main pulse peak height to the observations. These reductions may be attributed to the $\sim 2'$-off axis position of the pulsar during the observation, as well as the overestimate of \cite{Kuiper_Hermsen+10a}'s broad band pulse flux in the dominant $\sim$keV range (see their Fig.\,7). After the light curves were matched, we inspected the off-pulse maps, finding that the central pixels in the simulation were systematically too bright. This was addressed by simulating with additional blurring applied to our (on-axis) model PSFs. The best $\chi^2$ match of the simulated and data images used a $\sigma=12^{\prime\prime}$ Gaussian blur. \\
\begin{figure}
\centering
\includegraphics[width=0.99\linewidth]{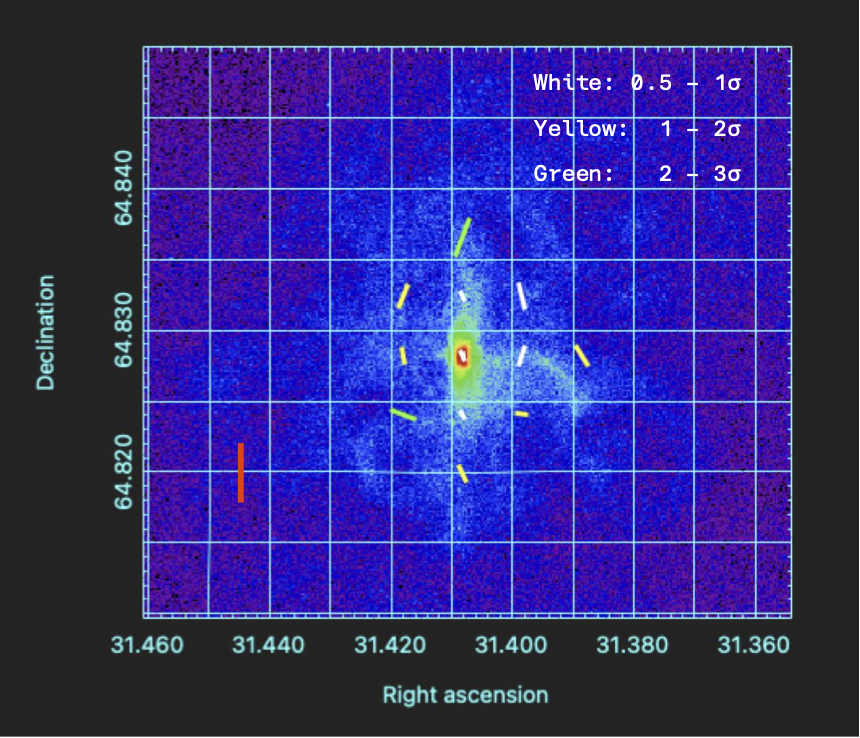}
\caption{A $5\times5$ $15''$ pixel grid of the nebula magnetic field (rotated $90^\circ$ from the EVPA) measured using simultaneous fitting. White bars denote $0.5-1\sigma$ significance, yellow $1-2\sigma$, and green $2-3\sigma$. The pixel field lies N-S, as expected for a toroidal magnetic field in the inner nebula. Other pixels show some EVPA variation, possibly following the filamentary structures of the background {\sl Chandra} image. A flux cut of ${>}700$ counts removes spurious polarization vectors in the outskirts. The red bar represents PD=100\%.}
\label{fig:nebula_polarization}
\end{figure}
\\
%%%%%%%%%%%%%%%%%%%%%%%%%%%%%%%%%%%%%
We divided the data into 3 phase bins representing the main pulse, interpulse, and off-pulse, modeled a single $2-8$ keV energy bin, and computed a $5 \times 5$ spatial grid of $15^{\prime\prime}$ pixels. The binning scheme was chosen to minimize the number of low-count bins ({$\sim$}12\% of these bins had $<10$ counts) while adequately resolving the PSF and phase structure. Polarization leakage correction was applied using the method described in \cite{Dinsmore24}. We obtained Q/I = $-0.20 \pm 0.57$ and U/I = $0.83 \pm 0.57$ for the pulsar main pulse and Q/I = $0.21 \pm 1.73$ and U/I = $-0.41 \pm 1.73$ for the pulsar interpulse. Fig. \ref{fig:nebula_polarization} shows the nebula polarization vectors (rotated by $90^\circ$ to show magnetic field direction) overlaid on the {\sl Chandra} image. Unfortunately, there were too few counts to allow a full resolution of the inner nebula and the pulsar did not produce significant polarization in either pulse phase bin. However, the lower significance magnetization vectors in the inner $\sim 30^{\prime\prime}$ do follow the expected N-S orientation of a central torus. The most significant pixels do show some PA variation, which may hint at more complex fields associated with the apparent jet and X-ray filamentary structure. Note that this analysis is consistent with the coarser spatial regions explored above: integrating over all the $15^{\prime\prime}$ pixels, we obtain PD$=18\% \pm 3.3\%$ and a $EVPA=99^\circ \pm 5^\circ$, consistent with the \texttt{XSPEC} and \texttt{3ML} measurements. \\

\subsection{Polarization modeling}
\label{sec:model}

\begin{figure}  %  Could be placed before the previous figure.
    \centering
    \includegraphics[width=0.99\linewidth]{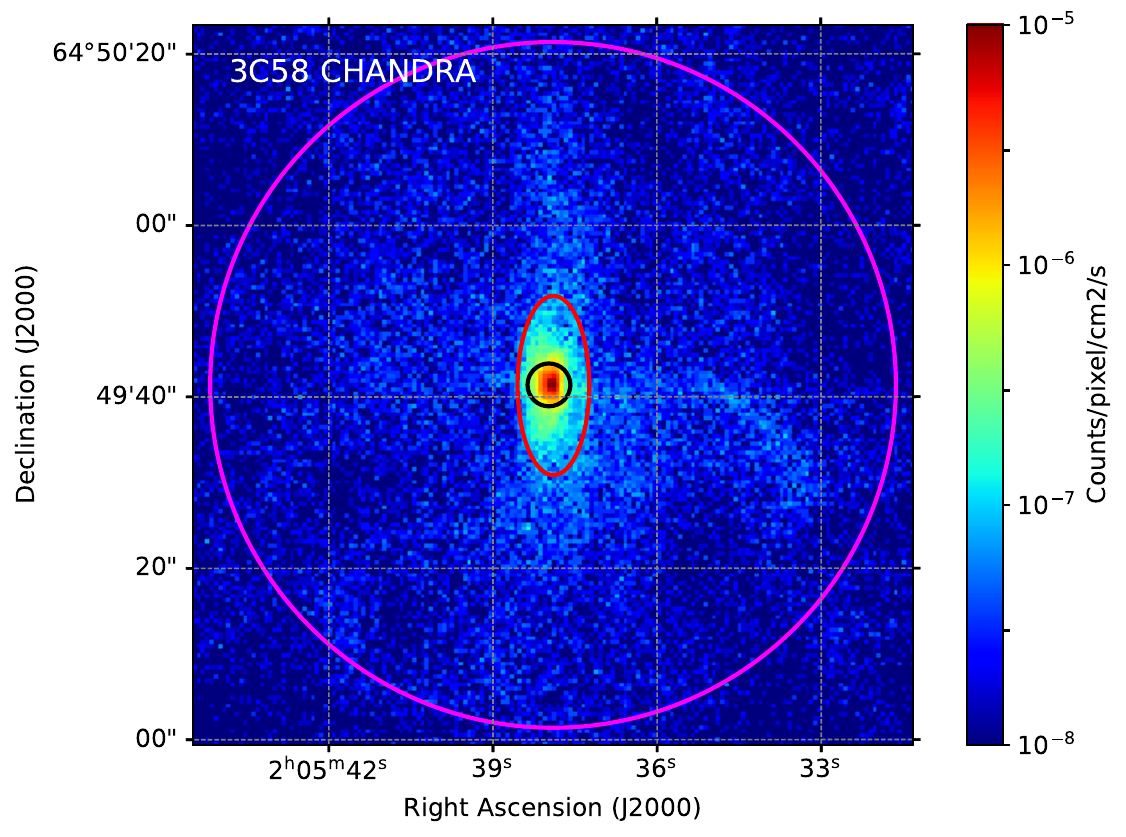}\\
    \caption{{\sl Chandra} count flux in the in the [2-8]~keV band of the inner PWN (same as Fig.~\ref{fig:counts}). The outer magenta circle corresponds to the 40~arcsec radius region used in the current analysis. The inner black circle identifies the region of the Inner Torus. The red ellipse identifies the region of the Extended Torus.}
    \label{fig:regions}
\end{figure}

In Fig.~\ref{fig:pol} the highly significant polarization of the central region is shown at top. In the lower panel, the points give the polarization degree computed for pulsar-centered circular regions of varying aperture, showing a decreasing trend in PD with region size. By performing a sample of mocked simulations of a uniformly polarized PWN, we determined that the observed trend is incompatible with a uniformly polarized PWN at about $1-1.5\sigma$ confidence level. To check compatibility with radio polarization maps, we performed an \texttt{ixpeobssim} simulation of this source, applying the radio polarization map of Sect.~\ref{sec:imaging} to the X-ray image as defined by the {\sl Chandra} events. Repeating the multi-radius aperture extraction, we observe in the simulated data (orange curve) a trend of decreasing polarization from PD$\simeq 8\%$ to $\simeq 4\%$, with $PA\simeq 90^\circ$. While much lower than the X-ray PD, the similar decrease in the radio suggests a common origin. To probe its nature, we consider four possible scenarios: (A) a uniformly polarized inner torus (the region within 2.5~arcsec from the PSR) embedded into an unpolarized nebula; (B) a uniformly polarized inner torus embedded into a nebula with the radio polarization imposed; (C) a uniformly polarized extended torus (an elliptical region with 10$"$ major axis) embedded in an unpolarized nebula; (D) a uniformly polarized extended torus embedded into a nebula with the radio polarization structure. See Fig.~\ref{fig:regions} for the region locations. We have ignored the (unpolarized) thermal X-ray emission, which, while $\sim 20$\% of the total PWN emission in the [2-6]~keV energy range \citep{Picquenot+24a}, is much less centrally peaked than non-thermal component, with an estimated contribution $<5$\% in the central 40" region.\\

In Fig.~\ref{fig:regions} we show  the two `torus' regions on the {\sl Chandra} image. The inner torus contains about 10\% of the total PWN counts, and about 25\% of the counts within a region of 40~arcsec radius. The PSR contribution to the emission within this same region is found, both from imaging and timing, to be approximately $\approx 10\%$, and is neglected in this analysis. The extended torus contains about 15\% of the total PWN counts, and about 40\% of the counts within a region of 40~arcsec radius. Both cases A and B require an inner torus with an intrinsic PD $ \approx 75\% \pm 5\%$ in order to match the high polarization found by IXPE. This is at the very limit of the theoretical maximum for synchrotron radiation ($\approx 70\%$) since some projection depolarization is expected, even for a near edge-on viewing angle. Given the weak (radio-matching) nebular polarization assumed here, the results are quite insensitive to the extended emission. The larger case C/D torus requires a more modest, but still large PD $\approx 55\% \pm 5\%$ and PD $\approx 50\% \pm 5\%$ respectively. This PD level is similar to that found previously in other spatially-resolved PWNe like Crab \citep{Bucciantini23a,Wong24a}, Vela \citep{Xie22}, and MSH~15-52 \citep{Romani23a}, supporting the idea that a larger region contributes to the central polarization. All models exhibit a steeper fall-off than the data. This could be explained if the data's extended nebula PD is about a factor 2-to-3 higher than the radio-set values applied here. 
% If I am interpreting what NB did correctly, this is all that this means. If so it wiuold be interesting to boost the PD of the extended emission by ~3x from the radio mp and see how the required torus PD and model fall-off rate change.
Some of the difference might also be attributed to the imperfectly calibrated {\it IXPE} response at this substantial off-axis angle. 
%This could be either attributed to the fact that \texttt{ixpeobssim} simulations assume the on-axis PSF, while this source was observed at about 2~arcmin off-axis, where the PSF is broader, or to a non perfect calibration of the \texttt{ixpeobssim}  PSF with respect to on-flight data \citep{Bucciantini23b,Dinsmore24}, even if we cannot at this time quantify these effects properly.

\section{Conclusions}
\label{sec:concl}
With the observation of \src, IXPE has covered nearly all young extended PWNe. Together with previous targets, these new data confirm a general trend suggesting a high level of polarization in the inner region of these nebulae, much higher than expectations based on either naive extrapolation from existing long wavelength data or on numerical simulations. In the case of \src\, we have verified that, under any reasonable assumption for the size of the polarized central region, the compact or extended torus, the inferred level of polarization implies a highly ordered magnetic field. This contrasts with recent PWN modeling that suggests a major role of turbulence in the acceleration region close to the pulsar termination shock \citep{Zhdankin17a,Bresci-Lemoine+23a}. However, we wish to remark that polarization is not a direct measure of magnetic field organization but, ultimately, a measure of turbulence isotropy. This implies that we cannot rule out the presence of a strongly anisotropic turbulence, an issue that would need further investigation both in terms of possible observable consequences, and theoretical implications. Unfortunately, with \src being relatively faint, our ability to tease out fine structure in this PWN with {\it IXPE} is limited and we cannot assess the possible patchiness of the PD. Similarly, with {\it IXPE}'s modest spatial resolution, the pulsed signal was not well separated from the surrounding nebula and we only derive weak upper limits on its polarization. With its modest 2-6\,keV flux, it does not significantly perturb the nebular analysis at our present sensitivity. While the integrated PD is very close to that of the Crab, our modeling suggests that the inner regions might potentially have a much higher PD and in this respect be closer to what is seen in Vela. If the age of 3C 58 is  older than what is derived from its traditionally claimed association with SN 1181, and closer to the characteristic age of PSR J0205+6449, this might imply a possible relation between turbulence in PWNe and their age, likely to be traced to the possible development of Rayleigh-Taylor instability which is known to be stronger in the earlier phases.  \\
\\
An interesting aspect of the current analysis is the high level of polarized background found in coincidence with a period of moderate solar activity, and the difficulty encountered in filtering out this contribution. We showed that the standard background rejection tools based on track properties do not adequately suppress this background, and that even removal of flaring epochs flagged by background rate increases does not completely remove this polarized flux. This deprecates the naive assumption that background can always be treated (and subtracted) as an unpolarized contribution. One should thus account for background polarization, especially in dealing with faint and extended sources during periods of high solar activity. 

\begin{acknowledgements}
 The Imaging X-ray Polarimetry Explorer (IXPE) is a joint US and Italian mission.  The US contribution is supported by the National Aeronautics and Space Administration (NASA) and led and managed by its Marshall Space Flight Center (MSFC), with industry partner Ball Aerospace (contract NNM15AA18C). This research was supported in part by grant 80NSSC25K7096. The Italian contribution is supported by the Italian Space Agency (Agenzia Spaziale Italiana, ASI) through contract ASI-OHBI-2022-13-I.0, agreements ASI-INAF-2022-19-HH.0 and ASI-INFN-2017.13-H0, and its Space Science Data Center (SSDC) with agreements ASI-INAF-2022-14-HH.0 and ASI-INFN 2021-43-HH.0, and by the Istituto Nazionale di Astrofisica (INAF) and the Istituto Nazionale di Fisica Nucleare (INFN) in Italy.  This research used data products provided by the IXPE Team (MSFC, SSDC, INAF, and INFN) and distributed with additional software tools by the High-Energy Astrophysics Science Archive Research Center (HEASARC), at NASA Goddard Space Flight Center (GSFC).  N.B. was supported by the INAF MiniGrant ``PWNnumpol - Numerical Studies of Pulsar Wind Nebulae in The Light of IXPE." F.X. is supported by National Natural Science Foundation of China (grant No. 12373041 and No. 12422306), and Bagui Scholars Program (XF). C.-Y.N. and S.Zhang are supported by a GRF grant of the Hong Kong Government under HKU 17304524.
\end{acknowledgements}

\bibliographystyle{aa}
\bibliography{3c58}
\end{document}